\documentclass[reprint, amsmath, amssymb, aps, pra]{revtex4-2}

\usepackage{subfigure}
\usepackage{amsmath}
\usepackage{physics}
\usepackage{enumitem}
\usepackage{graphicx}
\usepackage{bm}
\usepackage[colorlinks=true,allcolors=blue]{hyperref}
\usepackage{adjustbox}

\begin{document}
	
\title{Anisotropic Rabi model with two-photon relaxation}

\author{Hui Li}
\thanks{These authors contributed equally to this work.}
\author{Jia-Kai Shi}
\thanks{These authors contributed equally to this work.}
\author{Li-Bao Fan}

\author{Zi-Min Li}
\email{zimin.li@csu.edu.cn}

\author{Chuan-Cun Shu}
\email{cc.shu@csu.edu.cn}

\affiliation{
Institute of Quantum Physics, Hunan Key Laboratory of Nanophotonics and Devices, Hunan Key Laboratory of Super-Microstructure and Ultrafast Process, School of Physics, Central South University, Changsha 410083, China
}

\date{\today}

\begin{abstract}
{The interplay of three light-matter interaction processes - rotating and counter-rotating interactions and two-photon relaxation of the light field - is a topic of interest in quantum optics and quantum information processing. In this work, we theoretically investigate the three light-matter interaction processes using the anisotropic Rabi model, which accounts for different strengths of rotating and counter-rotating interactions and the unique occurrence of photon escape exclusively in pairs. By numerically solving the Lindblad master equation, we analyze the excitation-relaxation dynamics and derive a non-Hermitian effective Hamiltonian to gain further physical insights.
To explore the individual effects of these interactions, we examine three analytically tractable limits of the effective Hamiltonian. Our analysis reveals that the three competitive light-matter interaction processes exhibit sensitivity to parity, leading to intriguing phenomena in both transient and steady states. Particularly interesting dynamical patterns resembling quantum phase transitions emerge when these three interaction terms compete. This work deepens the understanding of ultrastrong light-matter interaction in open quantum systems and offers valuable insights into cavity-based quantum computations.}

\end{abstract}

\maketitle

\section{Introduction}

The quantum Rabi model (QRM), which characterizes the dynamic interplay between a two-level quantum system (referred to as a qubit) and a single-mode quantized light field confined within a cavity, stands as one of the most elementary interaction physical models \cite{Rabi_1936, Braak_2011, Xie_2017, Braak_2016, Zhong_2013, Zhong_2014}. Despite its fundamental nature, the QRM manifests a diverse array of physical phenomena and has discovered practical utility across an extensive spectrum of scientific disciplines. These include, but are not limited to, the realms of quantum optics \cite{Vedral_2005}, condensed matter physics \cite{Reik_1982, Wagner1986}, and the forefront of cutting-edge quantum technologies \cite{Yoshihara_2016, Frisk_Kockum_2019}.

The QRM features a competitive interplay between two distinct modes of light-matter interaction: the ``rotating terms" (RTs) and the ``counter-rotating terms" (CRTs). 
The RTs characterize processes involving the exchange of excitations between the qubit and the cavity, preserving the total excitation count. 
In contrast, the CRTs give rise to events where both a qubit excitation and a photon are simultaneously created or annihilated, maintaining only the parity of the excitation number.  The conservation of parity within the QRM leads to a multitude of significant physical consequences, including its solvability \cite{Braak_2011, Chen_2012, Zhong_2013, Li2021GAA, Li2021}, the emergence of level crossing points \cite{Li_2015, Batchelor_2015}, the presence of hidden symmetries \cite{Li2021a, Lu_2021, Mangazeev_2021}, and the occurrence of a super-radiant phase transition \cite{Hwang_2015, Liu2017, Chen_2020}.

While most research on the QRM has focused on idealized closed systems without relaxation, real-world situations involve environmental influences. 
Accounting for the environment requires the use of the Lindblad master equation. 
In certain cases, a non-Hermitian effective Hamiltonian can be derived to describe the system in the Schrödinger formalism, offering deeper physical insights. 
For example, an intriguing two-photon relaxation process has been postulated and investigated \cite{Wolinsky1988, Malekakhlagh_2019, Mirrahimi2014}. 
In this context, the escape of photons from a cavity occurs exclusively in pairs, preserving the conservation of parity within the QRM. 
As a consequence, the excitation-relaxation dynamics of the system, including both transient and steady states, exhibit a sensitivity to parity, giving rise to various remarkable phenomena that are absent in the closed QRM.
The two-photon relaxation mechanism has been proposed to facilitate universal quantum computing. 
Specifically, it has been theoretically and experimentally demonstrated that any initial state of the cavity, subjected to two-photon relaxation, evolves into two Schrödinger cat states, resulting in a qubit that relies on the characteristics of a cavity \cite{Mirrahimi2014, Leghtas2013, Leghtas2013a, Schuster2007, Leghtas2015, ofek2016, Wang2016}. 
The intricate interplay between the light-matter interaction and the two-photon relaxation has been explored in Ref.~\cite{Malekakhlagh_2019}. 
However, there is still a lack of comprehensive understanding regarding the competitive dynamics involving RTs, CRTs, and relaxation terms.

In this work, we adopt the theoretical framework established in Ref.~\cite{Malekakhlagh_2019} and investigate the competition among different interaction processes. 
Specifically, we consider the anisotropic Rabi model, which has been extensively studied in previous works \cite{Xie_2014, Tomka_2014, Zhang2015a, Joshi2016, Wang_2019, SanchezMunoz2020, Ying2021, Ying2022, Wang2023, Zhu2023a, Xu2024}.
In this model, the coupling strengths of the RTs and CRTs are different. 
By adjusting the coupling strengths and the two-photon relaxation rate, we find that the excitation-relaxation dynamics exhibit interesting parity-sensitive behavior in both transient and steady states. 
We explain these phenomena by analyzing the competition among different interaction processes. 
To achieve this, we numerically solve the quantum master equation and utilize the effective Hamiltonian to gain further physical insights.

The structure of this paper is as follows. 
In Section \ref{SectionModel}, we present the system Hamiltonian and the Lindblad master equation that incorporates two-photon relaxation. 
We subsequently explore three solvable limits in Section \ref{SectionSpectrum}, which are indicative of the effects arising from distinct transition processes. 
Section \ref{SectionDynamics} showcases exceptional phenomena observed in both transient dynamics and steady states. 
These are elucidated by examining the underlying principles of competitive interactions. 
Finally, a summary is given in Section \ref{SectionSummary}.

~

\section{Model Hamiltonian}\label{SectionModel}

We employ the anisotropic Rabi model (ARM) \cite{Tomka_2014, Xie_2014, Wang_2019, Zhang_2017} to examine the consequences of distinct light-matter interaction components. 
Different from dealing with the standard QRM, we encounter two types of interactions characterized by different strengths in the ARM. 
The ARM Hamiltonian reads 
\begin{equation}\label{ARM}
    \mathcal{H} = \mathcal{H}_0 + \mathcal{H}_{1} + \mathcal{H}_{2},
\end{equation}
where the free Hamiltonian $\mathcal{H}_0$, the Jaynes-Cummings interaction Hamiltonian $\mathcal{H}_{1}$ and the counter-rotating interaction Hamiltonian $\mathcal{H}_{2}$ are given by ($\hbar=1$)
\begin{equation}\label{ARMsplit}
    \begin{split}
        &\mathcal{H}_0 = \omega a^\dagger a + \frac{\Delta}{2}\sigma_z, \\ 
        &\mathcal{H}_{1} = g_1\left(a\sigma_+ + a^\dagger \sigma_- \right), \\
        &\mathcal{H}_{2} = g_2 \left(a^\dagger \sigma_+ + a \sigma_- \right). \\
    \end{split}
\end{equation}
Here, $a^\dagger$ and $a$ are the creation and annihilation operators for the quantized light field in the cavity, with a resonance frequency $\omega$.
Meanwhile, $\sigma_z = \ket{e}\bra{e} - \ket{g}\bra{g}$, $\sigma_+ = \ket{e}\bra{g}$, and $\sigma_- = \ket{g}\bra{e}$ are the atomic raising and lowering operators linked to the qubit with a transition frequency $\Delta$. 
The two types of interactions between the cavity and qubit are characterized by the coupling strengths $g_1$ and $g_2$. 
The ARM simplifies to the standard QRM with $g_1=g_2$. 
When $g_2=0$, it transforms into the Jaynes-Cummings model (JCM).  
When $g_1=0$, it corresponds to the anti-JCM (AJCM). 
Physical implementations or simulations of the QRM and ARM are achievable, e.g., in circuit quantum electrodynamics setups \cite{Yoshihara_2016, Wang_2019} and in synthetic antiferromagnets with intrinsic asymmetry of magnetic anisotropy \cite{Wang2024a}.

\begin{figure}[tb]
    \includegraphics[width=\linewidth]{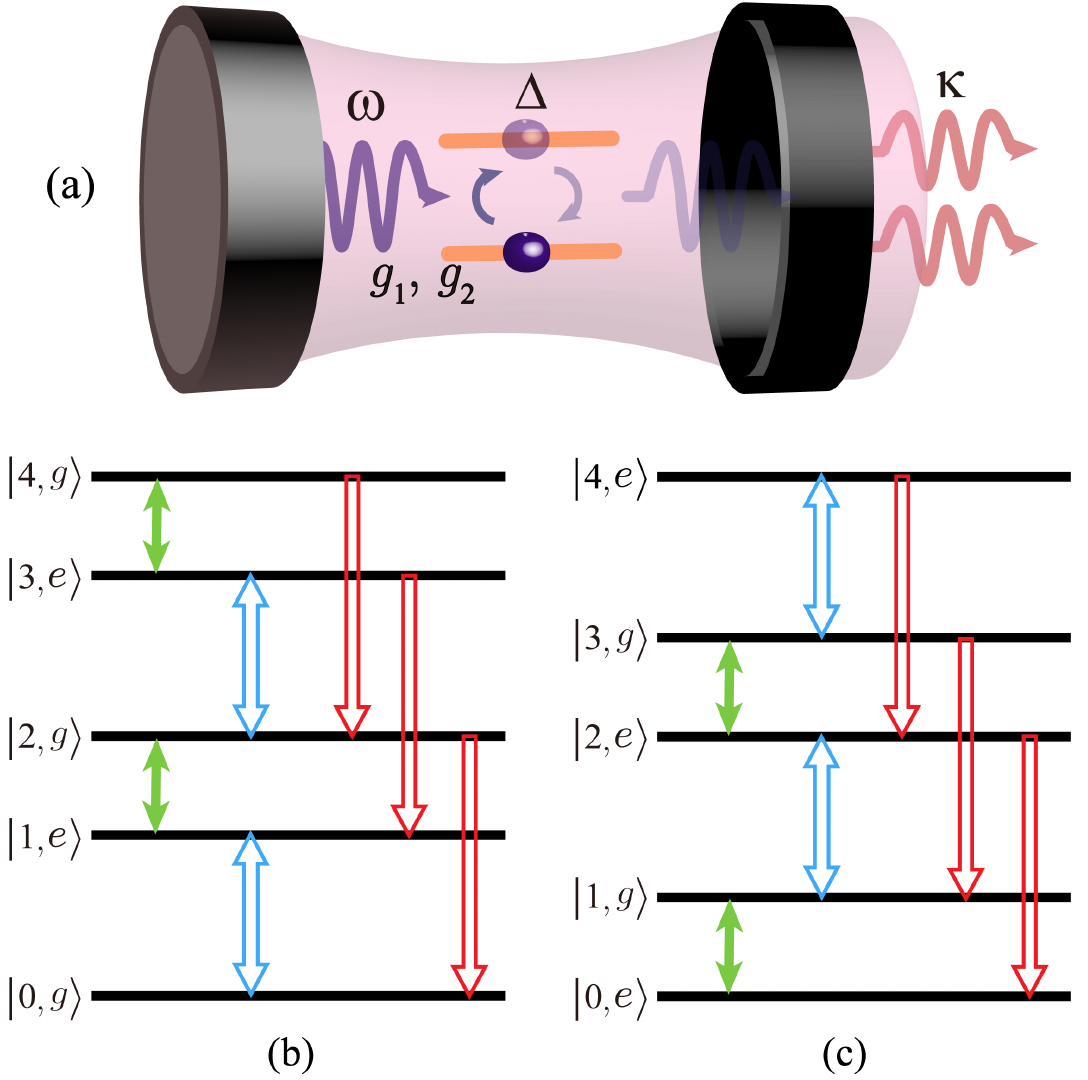}
    \caption{(a) A schematic of the anisotropic Rabi model with two-photon relaxation. (b) and (c) illustrate the possible transitions induced by the rotating interaction terms (green), counter-rotating terms (blue), and the two-photon relaxation term (red), in even and odd parity subspaces of photon-qubit bare basis, respectively. }
    \label{ARMtransitionFig}
\end{figure}

The Hamiltonian has the $Z_2$ symmetry in the sense that it is invariant under the parity transformation generated by the combined parity operator
\begin{equation}
    \mathcal{P} = -\sigma_z e^{i\pi a^\dagger a}.
\end{equation}
As a physical consequence, the ARM conserves the parity of the total excitation number in the system \cite{Li2021a}. 
To be specific, we consider the photon-qubit bare basis, defined by $\ket{n,e(g)}\equiv \ket{n}\otimes \ket{e(g)}$, with $\ket{n}$ being the Fock states, \(\ket{e}\) and \(\ket{g}\) being the excited and ground states of \(\sigma_z\).
Owing to the conserved parity, from an initial state with definite parity, the evolution under the ARM Hamiltonian goes along the corresponding parity chain \cite{Casanova_2010, Malekakhlagh_2019}, given by
\begin{equation}\label{paritychain}
    \begin{split}
        p=+1,\quad \left\{\ket{0,g}, \ket{1,e}, \ket{2,g}, \ket{3,e}, \dots \right\}, \\
        p=-1,\quad \left\{\ket{0,e}, \ket{1,g}, \ket{2,e}, \ket{3,g}, \dots \right\}. \\
    \end{split}
\end{equation}

Regarding the effects of the environment, a relaxation scheme where photons can only leak in pairs has been investigated \cite{Malekakhlagh_2019}.
In this case, the dissipation does not break the parity conservation, and the dynamical behaviors are expected to be parity-sensitive, i.e., different for initial states from two parity subspaces.  

When two-photon relaxation is considered, the time-dependent density matrix $\rho(t)$ of the system can be described by the Lindblad master equation \cite{wilma2018visualizing, wilma2019two, Manzano_2020}
\begin{equation}\label{MasterEquation}
    {\dot{\rho}(t)} = - i \left[\mathcal{H}, \rho(t)\right] +2\kappa a^2\rho(t)(a^\dagger)^2 - \kappa\{(a^\dagger)^2 a^2, \rho(t)\},  
\end{equation}
with $\kappa$ being the relaxation rate.

By rearranging Eq. (\ref{MasterEquation}), we obtain
\begin{equation}
    {\dot{\rho}(t)} = -i \left(\mathcal{H}_\text{eff}\rho(t) - \rho(t)\mathcal{H}_\text{eff}^\dagger\right) + \mathcal{L}_\text{jump}(\rho),
\end{equation}
with the effective Hamiltonian
\begin{equation}\label{Heff}
    \mathcal{H}_\text{eff} = \mathcal{H} - i\kappa (a^\dagger)^2 a^2,
\end{equation}
and the quantum jump operator
\begin{equation}
    \mathcal{L}_\text{jump}(\rho) = 2\kappa a^2\rho(t) \left(a^\dagger\right)^2.
\end{equation} 

When the quantum jump effects can be neglected, the dynamics of the system can be approximately described in Schr\"{o}dinger formalism with the effective Hamiltonian $\mathcal{H}_\text{eff}$.
The imaginary parts of eigenvalues of $\mathcal{H}_\text{eff}$ represent the decays of the eigenstates, and thus the leaks of probabilities from the system into the environment. 
Therefore, when calculating the expectation value of a physical observable in non-Hermitian systems, one needs to renormalize the wavefunction. 
Specifically, the imaginary term $i\kappa (a^\dagger)^2 a^2 $ in Eq.~(\ref{Heff}) indicates that higher photon levels have larger decay rates and thus decrease faster in probabilities. 
As a result, the states with slower photon decay become dominant. 
This is effectively a relaxation process with probabilities transiting from high photon levels to low photon levels, coinciding with the effects of the quantum jump operator $\mathcal{L}_\text{jump}(\rho)$. 
In this way, the TDSE of the effective Hamiltonian captures the main features of the dynamics apart from the quantum jump effects. 

Figure \ref{ARMtransitionFig} shows the transition diagram induced by the rotating interaction $\mathcal{H}_{1}$, the counter-rotating term $\mathcal{H}_{2}$ and the quantum jump operator $\mathcal{L}_\text{jump}(\rho)$. 
The rotating interactions (denoted with green arrows between $\ket{n,e}$ and $\ket{n+1,g}$) exchange excitations between the qubit and the cavity, and thus conserve the total number of excitations. 
The CRTs (indicated in blue arrows between $\ket{n,g}$ and $\ket{n+1,e}$) create (or annihilate) excitations in both the qubit and cavity simultaneously, and thus only conserve the parity of excitation numbers.
The quantum jump operator $\mathcal{L}_\text{jump}(\rho)$ (shown in red arrows) only happens from high levels to low levels. 
The decay rates are larger in the states with more photons. 
The interplay of these processes induces the exotic dynamical behaviors of the system. 

~

\section{Complex energy spectrum}\label{SectionSpectrum}

\subsection{Spectrum of the effective Hamiltonian}

\begin{figure}[tb]
    \centering
    \includegraphics[width=\linewidth]{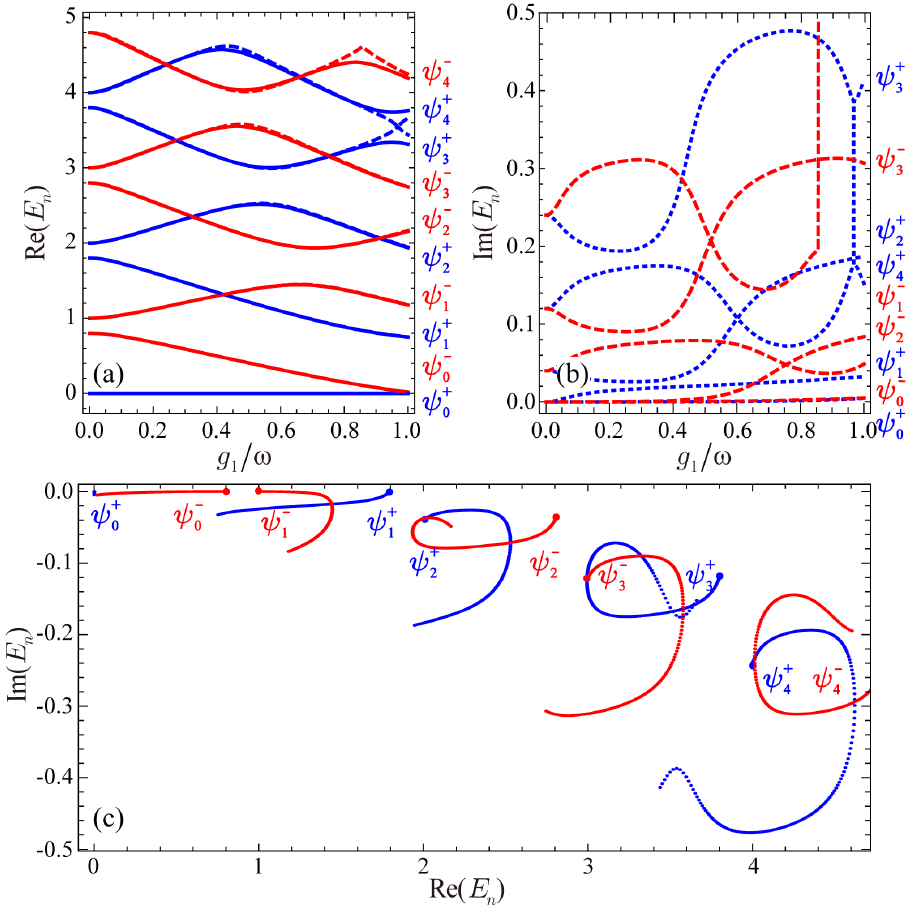}
    \caption{(a) Real and (b) imaginary parts of the eigenvalues of the ARM with two-photon relaxation, with respect to the coupling strength $g_1/\omega$. (c) Eigenvalues of the system Hamiltonian on the complex plane with various values of the coupling strength $g_1/\omega$. The fixed parameters are $\Delta/\omega=0.8$, $\kappa/\omega=0.02$, $\lambda=g_2/g_1=0.5$.  }
    \label{ARMSpectrumFig}
\end{figure}

As mentioned, the open system dynamics can be described by the Sch\"odinger equation of the effective non-Hermitian Hamiltonian.
In the following, we analyze the spectrum of this Hamiltonian.

There is no simple closed-form solution to the ARM and we thus rely on numerical diagonalization, as well as approximations. 
We first calculate the spectrum by numerical diagonalization and give physical interpretations to the complex eigenvalues.
The spectrum is complex as shown in Fig.~\ref{ARMSpectrumFig}.
The real parts of eigenvalues denote the frequencies and the imaginary parts characterize the effective decay rates.  
In Fig.~\ref{ARMSpectrumFig}, the eigenvalues are labeled with $\psi_n^p$, where $n$ is the level index and the value of $p$ takes $e$ for even parity subspace and $o$ for odd parity subspace. 

The non-Hermitian term $i\kappa a^\dagger a$ in the effective Hamiltonian does not directly induce any transition between bare states. 
However, it contributes imaginary eigenvalues associated with different photon numbers, and thus effectively results in decay processes. 
Therefore, the effective decay term in the Schr\"odinger formalism approximates the effects of the relaxation processes in the Lindblad formalism. 
In this way, we see intuitively how the non-Hermitian effective Hamiltonian can describe the full dynamics that are determined by the quantum master equation.\\ \indent
We turn to solvable limits of the effective Hamiltonian for further physical intuitions. The JCM and AJCM are both exactly solvable and thus the effects of the RTs and CRTs can be analyzed separately. 
From the analytic results, we can deduce the competition between different types of interactions. 
Even the trivial decoupled limit, with $g_1=g_2=0$, provides additional intuition to the decay process.  

\subsection{Decoupled limit}
With $g_1=g_2=0$, the qubit is decoupled from the field, and the effective Hamiltonian becomes
\begin{equation}
    \mathcal{H}_\text{eff}^{g=0} = \mathcal{H}_0 - i\kappa (a^\dagger)^2 a^2.
\end{equation}
This case is trivially solvable with the eigenvalues given as
\begin{equation}
    E^{g=0}_{n,\pm} = n\omega \pm \frac{\Delta}{2}- i\kappa n(n-1),
\end{equation}
and the eigenstates are simply bare states
\begin{equation}
    \psi^{g=0}_{n,\pm} = \ket{n,\pm} \equiv \ket{n}\otimes\ket{\pm}. 
\end{equation}

We see from the eigenvalues that the decay rates of eigenstates increase with the photon number. 

\subsection{Jaynes-Cummings model}
When $g_2$ is set to 0 and only rotating interactions are taken into account, we reach the  JCM \cite{Jaynes_1963, Fan2020, Fan2023PRL, fan2023pulse}. 
The effective Hamiltonian now reads
\begin{equation}
    \mathcal{H}_\text{eff}^\text{JC} = \mathcal{H}_0 + g_1\left(a\sigma_+ + a^\dagger \sigma_- \right) - i\kappa (a^\dagger)^2 a^2,
\end{equation}
which is exactly solvable. 

The eigenvalues are
\begin{equation}
        E_{n,\pm}^\text{JC} = \left(n+\frac{1}{2}\right)\omega - i\kappa n^2 \pm \frac{1}{2}\sqrt{(A_n^-)^2+B_n^2}
\end{equation}
where
\begin{equation}\label{AnBn}
    \begin{split}
        &A_n^\pm =  (\omega \pm \Delta) -2i\kappa n, \\
        &B_n = 2g\sqrt{n+1}.
    \end{split}
\end{equation}
The definition of $A_n^+$ will be used for the anti-Jaynes-Cummings model later. 

A special case that can be analytically dealt with is when the qubit and cavity are in resonance, with $\delta=\omega-\Delta=0$. 
The corresponding eigenvalues are
\begin{equation}\label{EnJC}
    E_{n,\pm}^\text{JC} = \left(n+\frac{1}{2}\right)\omega - i\kappa n^2 \pm \sqrt{(n+1)g_1^2 - n^2\kappa^2}.
\end{equation}
The two eigenstates with the same index $n$ are degenerate when the square root in Eq.~(\ref{EnJC}) is 0, leading to an exceptional point (EP) due to the passive $\mathcal{PT}$-symmetry, see further discussion in the Appendix \ref{AppendixPT}.

\subsection{Anti-Jaynes-Cummings model}
We now consider the anti-JCM to explore the effects brought by CRTs. 
The effective Hamiltonian now reads
\begin{equation}
    \mathcal{H}_\text{eff}^\text{AJC} = \mathcal{H}_0 + g_2 \left(a^\dagger \sigma_+ + a \sigma_- \right) - i\kappa (a^\dagger)^2 a^2,
\end{equation}
which is also exactly solvable.

The corresponding eigenvalues are
\begin{equation}
        E_{n,\pm}^\text{AJC} = \left(n+\frac{1}{2}\right)\omega - i\kappa n^2 \pm \frac{1}{2}\sqrt{(A_n^+)^2+B_n^2},
\end{equation}
with $A_n^+$ and $B_n$ given in Eq.~(\ref{AnBn}). 

~

\section{Excitation and relaxation dynamics}\label{SectionDynamics}

In this section, we analyze the dynamics of the ARM with two-photon relaxation. 
To this end, we tune the coupling strengths $g_1$ and $g_2$, as well as the two-photon relaxation rate $\kappa$, to explore the interplay and competition of the different interaction terms. 

We focus on both transient states and steady states by calculating the corresponding physical observables, namely the cavity photon number and the qubit population. 
In particular, we discuss the excitation-relaxation dynamics with three approaches. 
Firstly, we calculate the dynamics by numerically solving the Lindblad master equation with a specific initial state $\rho(0)$. 
We illustrate the overall dynamics in two parity subspaces and discuss the transient and steady states in detail. 
Secondly, we map the bare states onto the eigenbasis of the effective Hamiltonian and understand the relaxation dynamics by interpreting the corresponding complex eigenvalues as decay rates.
Thirdly, we understand some interesting steady-state behaviors from the perspective of the transition processes of the bare states. 

\subsection{Dynamical effects of the interaction terms}
To begin with, we intuitively analyze the impact of the three types of interactions on system dynamics. 
The relaxation processes involved in Eq. (\ref{MasterEquation}) is the simplest case as it does not include the interaction with the qubit. 
In addition, there is no process to regenerate photons into the system after the dissipation. 
In the decoupled limit, if we start from a bare state $\ket{n,e(g)}$, photons will decay at an exponential rate given by $\kappa n(n-1)$ and the cavity reaches the vacuum state $\ket{0,e(g)}$ if $n$ is even. 
For initial states with odd photons, however, there will be a single photon left in the system since there is no mechanism to dissipate it. 
Since the qubit and cavity are decoupled in this limit, the qubit state is immune to relaxation and remains unchanged. 

As shown in Fig.~\ref{ARMtransitionFig}, the JC interaction terms generate transitions within subspaces that conserve excitation numbers.
This leads to Rabi oscillations between two basis states that share the same number of excitations. 
Particularly, if we start with the initial state $\ket{n,e}$, the probability of state $\ket{n+1,g}$ is oscillatory, given by
\begin{equation}\label{JCoscillation}
    P^\text{JC}_n = \frac{g_1^2\sin^2(\Omega_n^\text{JC}t) }{(\omega-\Delta)^2+g_1^2},
\end{equation}
with the Rabi frequencies 
\begin{equation}\label{JCfrequency}
    \Omega_n^\text{JC} =  \sqrt{(\omega-\Delta)^2 + 4g_1^2(n+1)}.
\end{equation}
From Eq.~(\ref{JCoscillation}), we see that the complete population transfer only occurs in the resonant case. 
The oscillation frequencies increase with the coupling strength \(g_1\) and the photon number \(n\).

\begin{figure*}[t]
    \centering
    \includegraphics[width=.9\linewidth]{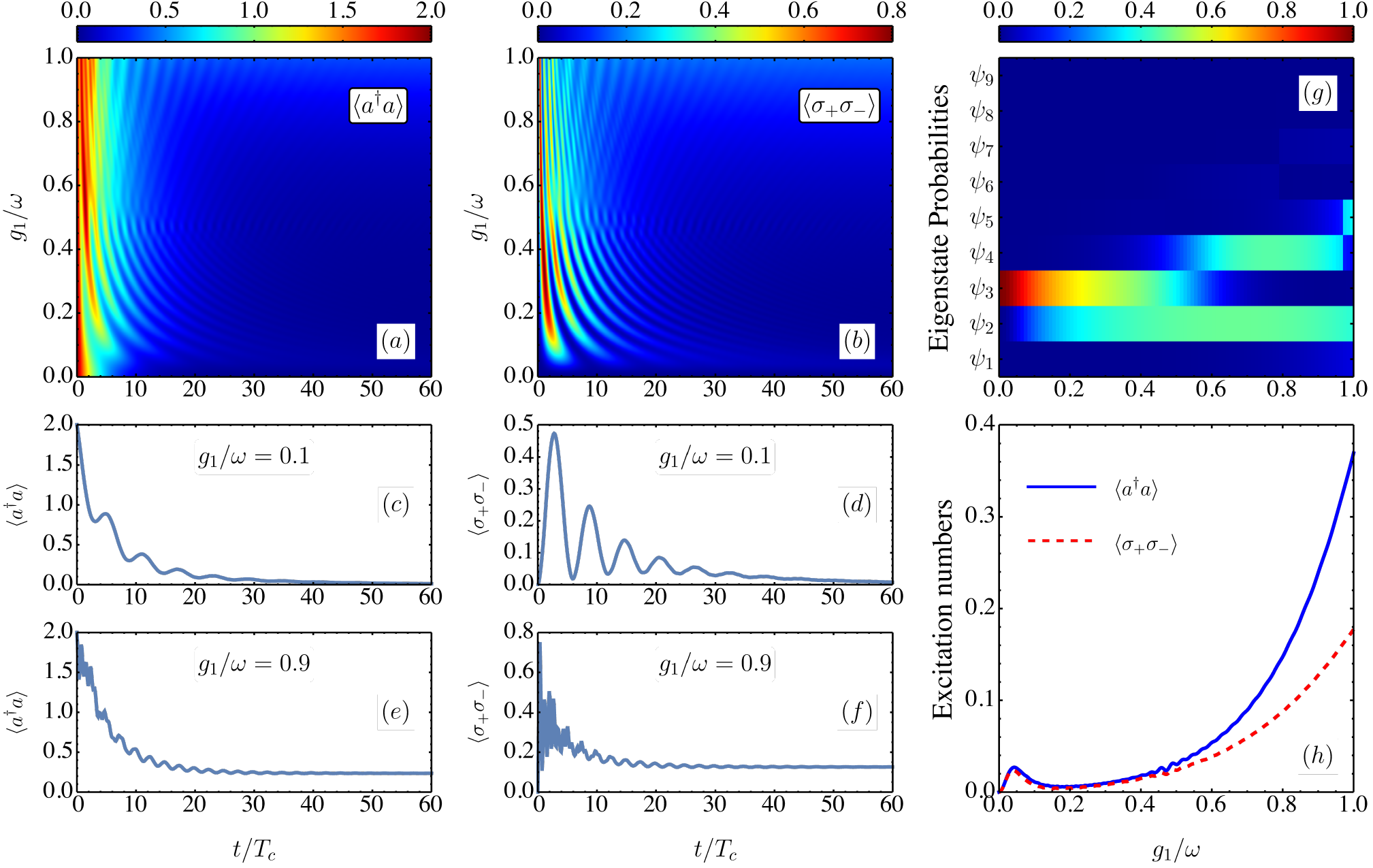}
    \caption{
        Excitation-relaxation dynamics of the ARM in even parity subspace, with the initial state $\ket{2,g}$, and parameters $\lambda=g_2/g_1 = 0.5$, $\Delta/\omega=0.8$, $\kappa/\omega = 0.02$. 
        Time evolution of (a)(c)(e) mean photon number and (b)(d)(f) qubit population with various values of coupling strength $g_1/\omega$ in the time range of $t/T_c$, with $T_c\equiv\pi/\omega$. 
        (g) The probability of eigenstates when mapping the initial state onto the eigenbasis. 
        (h) The transient states of mean photon number and qubit population with various values of coupling strength $g_1/\omega$ at time $t/T_c = 60$. 
    }
    \label{ARMDynamicsFig}
\end{figure*}

Different from the JC interaction terms, the CRTs induce transitions where two excitations emerge or disappear simultaneously, leading to anti-Rabi oscillations.
With the initial state being $\ket{n,g}$, the population of the state $\ket{n+1,e}$ is given by
\begin{equation}\label{AJCoscillation}
    P^\text{AJC}_n = \frac{g_2^2\sin^2(\Omega_n^\text{AJC}t) }{(\omega+\Delta)^2+g_2^2},
\end{equation}
with the anti-Rabi frequencies
\begin{equation}\label{AJCfrequency}
    \Omega_n^\text{AJC} = \sqrt{(\omega+\Delta)^2 + 4g_2^2(n+1)}.
\end{equation}
We can observe from the above Equations that for small values of \(g_2\), the frequencies are large while the amplitudes are small. 
Therefore, the effects of CRTs are negligible for small $g_2$ while prominent for large $g_2$. 

The intricate interplay of the above dynamical processes leads to the interesting time evolution of the full system.

\subsection{Transient states}

We now start our discussions on the dynamics with the transient states in the even parity subspace. 
The full dynamics with even parity are presented in Fig.~\ref{ARMDynamicsFig}.
The system is initially prepared in the bare state $\ket{2,g}$, indicating that there are two photons in the cavity and the qubit is in its ground state at $t=0$. 
We calculate the mean photon number and qubit population by solving the Lindblad master equation numerically. 
In the calculations, we set $g_2/g_1 = \lambda$ and vary the value of $g_1$ to obtain the dynamics concerning coupling strengths. 
To better see how RTs and CRTs interact with the relaxation separately, we have chosen the coupling ratio \(\lambda=0.5\) to further suppress the effects of CRTs for lower coupling strength regimes. 
The overall results are displayed in Fig.~\ref{ARMDynamicsFig}(a) and \ref{ARMDynamicsFig}(b).
It is clear to observe that the interplay of the three types of interactions depends on the coupling strengths. 
The dynamical behaviors can be physically interpreted from the perspective of transition processes induced by the three types of interactions, as illustrated in Fig.~\ref{ARMtransitionFig}(a).

We take the case \(g_1/\omega=0.1\) as the first example, with the corresponding cavity photon number and qubit population displayed in Fig.~\ref{ARMDynamicsFig}(c) and (d), respectively. 
The ``ripples" for short-range time mainly result from the interplay between the JC interactions and the two-photon relaxation, since the effects of CRTs are weak with the marginal value of \(g_2/\omega=0.05\). 
Through the relaxation channel \(\ket{2,g}\Rightarrow\ket{0,g}\), the two photons in the initial state quickly escape from the cavity with a certain probability.
Meanwhile, the Rabi oscillations transfer some populations through the process \(\ket{2,g}\leftrightarrow\ket{1,e}\) and thus prevent the relaxation. 
The effect of Rabi oscillations can be more obviously observed from the qubit dynamics in Fig.~\ref{ARMDynamicsFig}(d), where the qubit is not directly affected by the relaxation process and undergoes several periods of oscillations at the constant Rabi frequency.  
Later on, since the Rabi oscillations are reversible, the re-excited populations undergo the relaxation again, which gives rise to an overall decaying behavior for both cavity and qubit. 
Consequently, after some periods of Rabi oscillations, the populations concentrate in the state \(\ket{0,g}\), with both cavity and qubit sitting around their ground states. 

When $g_1$ increases, the effects of CRTs become non-negligible, and the interference between these effects leads to non-trivial dynamical patterns. 
To demonstrate this, we take the case \(g_1/\omega=0.9\), and thus \(g_2/\omega=0.45\), as an example.
The results are displayed in Fig.~\ref{ARMDynamicsFig}(e) and (f).
The current case has two main differences from the previous one: the photons decay slower at the initial stage, and more photons are left in the cavity in the final state.
Since CRTs induce fast small-amplitude oscillations, the dynamics at beginning times are largely different from the previous case. 
For this time range, the main transition channel owing to the competition between CRTs and relaxation is
\begin{equation}        
    \ket{2,g}\Leftrightarrow\ket{3,e}\Rightarrow\ket{1,e},
\end{equation}
where the relaxation is weakened to a much greater extent.
As a consequence, the mean photon number and qubit population in this case are both more than in the previous case at the beginning. 
After this initial competition, the reversible nature of Rabi and anti-Rabi oscillations still offers chances to the relaxation channel and thus leads to the overall decaying behavior. 
The relaxation processes smooth these oscillatory dynamics for a long time range, where all the transitions reach a dynamical equilibrium, with a higher number of excitations in the final state. 

We note that any other initial states with even parity give rise to almost the same physics since they decay quickly to the bare state $\ket{2,g}$ that we consider in the above example. 
However, higher initial states will lead to more complicated dynamics at the beginning times, since the Rabi and anti-Rabi frequencies increase with photon number \(n\).
The final dynamical equilibrium only depends on the system parameters and the parity of the initial state, ignoring the specific form of the wavefunctions.

If we focus on the medium-range time that is long (e.g., $t=60\pi$ in our case), but not long enough for the system to reach the final steady state, we can see how the equilibrium is affected by the coupling strengths. 
For the even parity subspace, we observe some dynamical behaviors that are rather steady, as illustrated in Fig.~\ref{ARMDynamicsFig}(d).
These medium-range states are already quite similar to the final steady states except for the local maximum at around $g_1=0.05$.
This peak was noticed in Ref.~\cite{Malekakhlagh_2019} and was believed to exist in the steady state for the coupling strength $g_1=g_2=\kappa$.

\begin{figure*}[t]
    \centering
    \includegraphics[width=.9\linewidth]{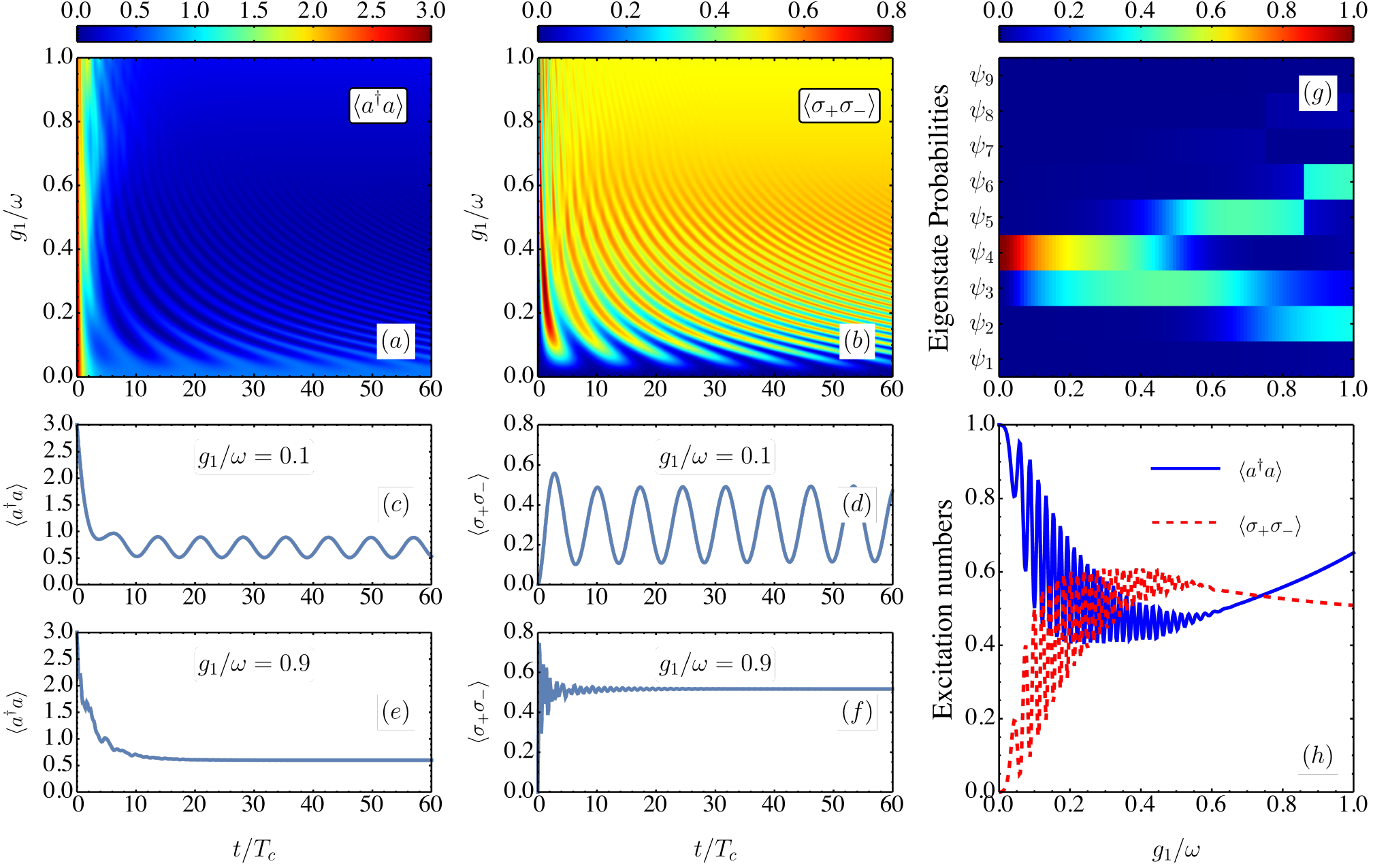}
    \caption{
        Excitation-relaxation dynamics of the ARM in odd parity subspace, with the initial state $\ket{3,g}$, and parameters $\lambda=g_2/g_1 = 0.5$, $\Delta/\omega=0.8$, $\kappa/\omega = 0.02$. 
        Time evolution of (a)(c)(e) mean photon number and (b)(d)(f) qubit population with various values of coupling strength $g_1/\omega$ in the time range of $t/T_c$, with $T_c\equiv\pi/\omega$. 
        (g) The probability of eigenstates when mapping the initial state onto the eigenbasis. 
        (h) The transient states of mean photon number and qubit population with various values of coupling strength $g_1/\omega$ at time $t/T_c = 60$. 
    }
    \label{ARMDynamicsFigOdd}
\end{figure*}

Here, we show that this local maximum is absent in the final steady states and its position is irrelevant to the relaxation rate $\kappa$. 
It turns out to be a consequence of the competition between JC interactions and two-photon relaxation, which is most clearly seen in the dissipative JCM, where CRTs are absent. 
In a more direct interpretation, the local maximum originates from the slow Rabi oscillations for small $g_1$, whose position can be analytically determined from the JCM. 
We see from Eq.~(\ref{JCfrequency}) that the Rabi frequency increases with the coupling strength \(g_1\). 
Therefore, for a given time \(t\), oscillations associated with different coupling strengths are going through different stages. 
There might be one that happens to be at its maximum, whereas the neighboring ones are not. 
At this specific time, we then observe the local maximum with respect to the coupling strength, as displayed in Fig.~\ref{ARMDynamicsFigOdd}(h).
With this reasoning, it is clear that the corresponding $g_1$ value decreases with time $t$, and thus the peak disappears for a long enough time. 

Since the total system conserves the parity of the total excitation number, it is expected that the dynamics are parity-sensitive. 
Indeed, the results with the initial odd-parity state $\ket{3,g}$, shown in Fig.~\ref{ARMDynamicsFigOdd}, demonstrate that the dynamics of both cavity photon number and qubit population are significantly different from the even parity case. 
The major difference is that the oscillations are much more persistent in odd subspace. 
The dynamics can also be understood in the same intuitive way and can be largely deduced from the solvable limits. 
In the decoupled case where $g_1 = g_2 =0$, the initial state $\ket{3,g}$ quickly loses two photons and reaches the final state $\ket{1,g}$. 
In this case, there is no mechanism to excite the qubit or dissipate the photon. 
Things become more interesting when light-matter interactions are introduced. 
For small coupling strength, the system oscillates between states $\ket{1,g}$ and $\ket{0,e}$ owing to the JC interactions. 
If $g_2=0$, this oscillation persists forever and the steady state is never reached. 
This is verified by the constant oscillations depicted in Figs.~\ref{ARMDynamicsFigOdd}(c) and \ref{ARMDynamicsFigOdd}(d).
For non-zero $g_2$, however, the weak effects of CRTs induce a tiny amount of transitions to higher photon states which are then dissipated by the relaxation processes. 
This is the reason why the system can still reach steady states after long-time oscillations. 
When $g_2$ grows to the regimes where the effects of CRTs are prominent, the Rabi oscillations are destructed by the CRTs and then quickly smoothed by the relaxation terms, as shown in Figs.~\ref{ARMDynamicsFigOdd}(e) and \ref{ARMDynamicsFigOdd}(f).
It is clear from the quasi-steady states in Fig.~\ref{ARMDynamicsFigOdd}(h) that the photon and qubit excitation left in the system is more than those of the even parity case. 
This is because there are always some populations of \(\ket{1,g}\) and \(\ket{0,e}\) that can not be dissipated.

Since the connections between the Lindblad master equation and the non-Hermitian effective Hamiltonians have been established in the preceding sections, the overall dynamics can also be understood with physical intuitions by mapping the initial state on the eigenbasis of the Hamiltonian (\ref{ARM}). 
This mapping shows which modes are more active at a given value of $g$ in each parity subspace. 
By doing so, we can check the corresponding frequencies and decay rates from Fig.~\ref{ARMSpectrumFig} and thus understand the dynamical behavior. 
Here, we map the initial bare state $\ket{2,g}$ onto the eigenstates of the effective Hamiltonian, and the probabilities of the eigenstates are displayed in Fig.~\ref{ARMDynamicsFig}(g).

\subsection{Steady states}

\begin{figure}[tb]
    \centering
    \includegraphics[width=\linewidth]{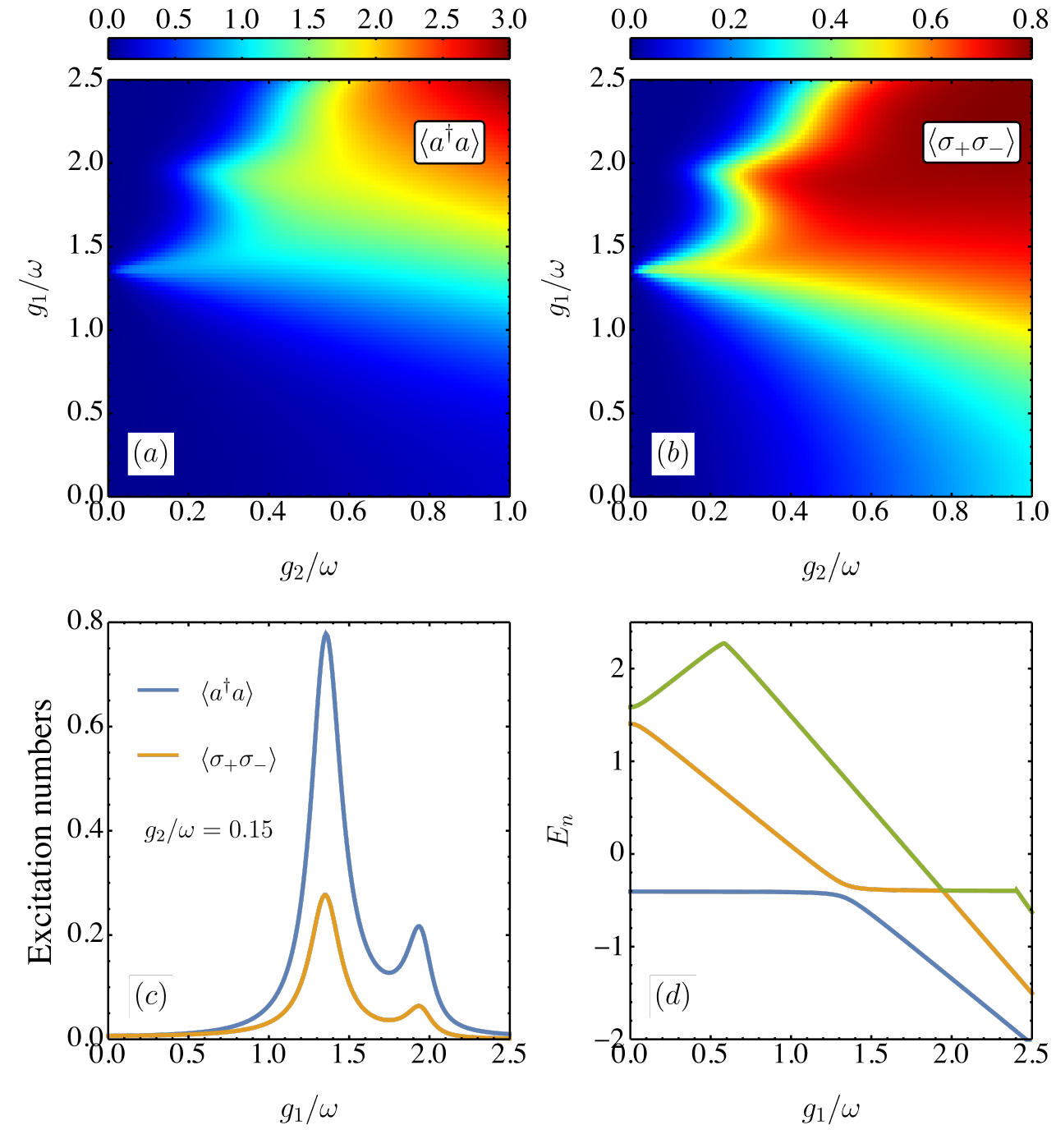}
    \caption{
    (a) The cavity photon number and (b) the qubit population in the steady state of even parity subspace with respect to the coupling strengths $g_1$ and $g_2$. 
    (c) The cavity photon number and the qubit population with fixed coupling $g_2/\omega=0.15$ and varying coupling strength $g_1/omega$. 
    (d) The eigenvalue spectrum with the same parameter values as in (c). The peaks in (c) coincide with the avoided crossings in (d). 
    }
    \label{SteadyStateEvenFig}
\end{figure}

\begin{figure}[tb]
    \centering
    \includegraphics[width=\linewidth]{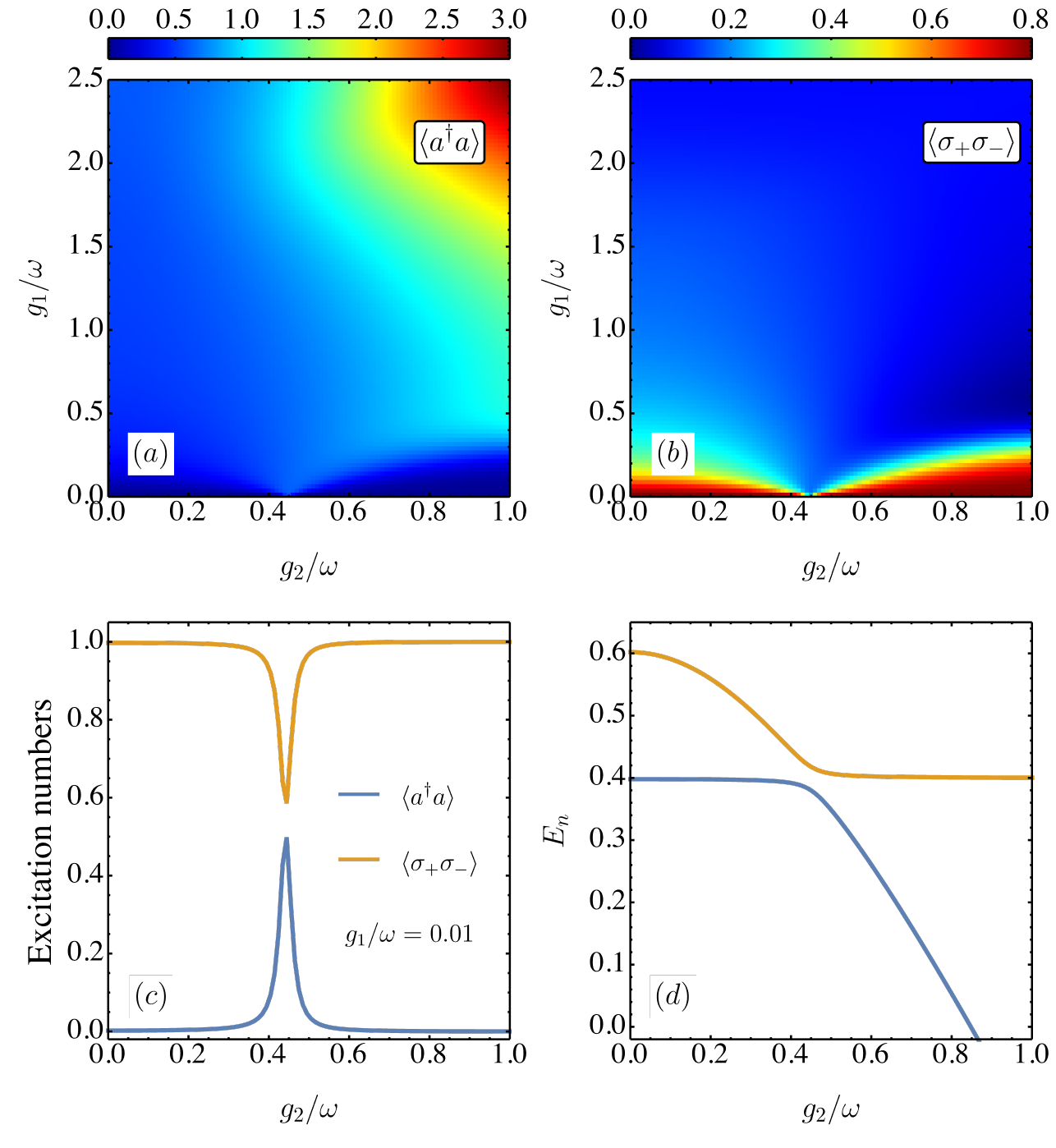}
    \caption{
        (a) The cavity photon number and (b) the qubit population in the steady state of odd parity subspace with respect to the coupling strengths $g_1$ and $g_2$. 
        (c) The cavity photon number and the qubit population with fixed coupling $g_1/\omega=0.01$ and varying coupling strength $g_2/omega$. 
        (d) The eigenvalue spectrum with the same parameter values as in (c). The peak and valley in (c) coincide with the avoided crossing in (d). 
    }
    \label{SteadyStateOddFig}
\end{figure}

After a long enough time, the system arrives at a dynamical equilibrium of the three types of transitions, i.e. the steady state.
With steady states, it is easy to identify the interplay of the interaction processes. 
The steady states for the cavity and the qubit in even and odd parity subspaces are calculated with respect to the coupling strengths $g_1$ and $g_2$ and displayed in Figs.~\ref{SteadyStateEvenFig} and \ref{SteadyStateOddFig}, respectively.
We can observe different behaviors by following different paths on the $g_1$-$g_2$ plane. 
For example, following the path $g_2=0$ we observe the steady states of the JCM with respect to $g_1$. 
On the other hand, we have the steady states of the AJCM with the constraint $g_1=0$. 
Setting $g_1=g_2$ we have the standard QRM, which exhibits parabolic behaviors with respect to the coupling strengths. 
We note that the steady states are determined solely by the parity, rather than any specific forms, of the initial states.
Nevertheless, for the sake of clarity and simplicity, we base our discussions on the particular initial states $\ket{2,g}$ for even parity and $\ket{3,g}$ for odd parity in the following.

To gain more physical intuitions, we again turn to the three solvable limits and study the interplay of coupling and relaxation terms. 
For the even parity case, the decoupled limit leads to the trivial consequence that both photons escape from the system through the relaxation $\ket{2,g}\Rightarrow\ket{0,g}$.
Therefore, there is no photon in the cavity after a long enough time, and the qubit stays in the ground state. 
In the JCM limit where $g_2=0$, the different situation gives rise to the same result.
As can be seen from the transient states of the JCM, the Rabi oscillation between $\ket{2,g} \leftrightarrow \ket{1,e}$ diminishes with time because the total populations decay exponentially to the vacuum bare state through the irreversible relaxation process $\ket{2,g} \Rightarrow \ket{0,g}$. 
With this, it is again verified that the small peak in Fig.~\ref{ARMDynamicsFig}(h) is not a steady state, but rather a ``quasi-steady" one. 

Following another path $g_1=0$, the dynamics can be described by the AJCM, in which case the transition processes are more complicated. 
The anti-Rabi oscillation $\ket{2,g}\Leftrightarrow \ket{3,e}$ competes with the relaxation, leading to the lower level anti-Rabi oscillation \(\ket{0,g}\Leftrightarrow\ket{1,e}\) that is immune to the further relaxation process. 
The amount of populations that are transferred by the anti-Rabi oscillation is determined by the \(P_1^\text{AJC}\) in Eq.~(\ref{AJCoscillation}). 
Consequently, the excitation numbers in the AJCM case increase with the value of \(g_2\). 
This anti-Rabi oscillation persists forever unless JC interactions associated with \(g_1\) are introduced. 

For the general case, with all three types of interactions present, the steady state of the ARM is a non-trivial mix of JCM and AJCM with two-photon relaxation. 
Generally, the steady states of cavity and qubit increase with respect to both \(g_1\) and \(g_2\), as can be observed from Figs.~\ref{SteadyStateEvenFig}(a) and \ref{SteadyStateEvenFig}(b) for cavity and qubit, respectively. 
The results along the path \(g_1=g_2\) are consistent with the results presented in Ref.~\cite{Malekakhlagh_2019}.
A particularly interesting phenomenon, absent in the standard QRM, of the steady states in even parity subspace is the peaks along $g_1$ with a small value of $g_2$, as shown in Fig.~\ref{SteadyStateEvenFig}(c). 
These peaks are found sitting at the avoided level crossings of the eigenspectrum, as displayed in Fig.~\ref{SteadyStateEvenFig}(d).
These avoided crossings can be analytically determined by calculating genuine level crossings of the JCM. 
The position for the \(m\)-th peak, with respect to \(g_1\), is then given by
\begin{equation}
    \frac{g_1}{\omega} = \sqrt{(2m-1)+\frac{\Delta}{\omega}},\quad m\in\mathbb{N}_+.
\end{equation}

To understand the physical origins of these peaks, we go back to the Hermitian case.
We start with JCM with \(U(1)\) symmetry.
The energy levels in even parity subspace cross with each other, since they now all belong to different symmetric sectors. 
If we add perturbative \(g_2\) into the system.
The \(U(1)\) symmetry is broken and reduced to the \(\mathbb{Z}_2\) symmetry. 
The spectrum with a specific parity subspace will repulse with each other, and thus the previous level crossings become avoided. 
At avoided crossings induced by small \(g_2\), eigenstates strongly mixed and thus the corresponding physical observables drastically change. 
In JCM, the initial state \(\ket{0,g}\) is the ground state living in a 1-dimensional subspace and it does not interact with any other eigenstates.
When we switch on the CRTs with small \(g_2\), the initial state \(\ket{0,g}\) strongly interacts with other states at the avoided crossings. 
Since the initial state \(\ket{0,g}\) is no longer the eigenstate of the system, the dynamics exhibit Rabi oscillation between the two original eigenstates at the avoided crossings. 
In the vicinity of the first peak at \(g_1/\omega=\sqrt{1+\Delta/\omega}\), the following oscillation occur,
\begin{equation}\label{EvenSSoscillation}
    \ket{0,g}\leftrightarrow \ket{1,-}_\text{JC}.
\end{equation}
The JC eigenstate (or dressed state) \(\ket{1,-}_\text{JC}\), explicitly given in Eq.~(\ref{JCeigenstates}), is a superposition of the bare states \(\ket{0,e}\) and \(\ket{1,g}\), and contains 1 excitation.
Correspondingly, the photon number and qubit population oscillate between 0 and 1. 
With the relaxation process considered, the oscillations will be smoothed and arrive at steady states with larger mean photon numbers. 
In the meantime, for other values of \(g_1\) that are far from the avoided crossings, the oscillation in Eq.~(\ref{EvenSSoscillation}) is weak and the final excitation is marginal.
If the initial state is not \(\ket{0,g}\), the JC interaction and relaxation will make the system jump to the lowest levels and thus lead to the same results as the initial \(\ket{0,g}\).

As expected, the steady states in the odd parity subspace are shapely different, as shown in Fig.~\ref{SteadyStateOddFig}. 
In the trivial decoupled limit, the initial state $\ket{3,g}$ irreversibly decays to the final state $\ket{1,g}$. 
In the JCM limit, the Rabi oscillation $\ket{1,g}\leftrightarrow\ket{0,e}$ persists forever, giving rise to the total excitation being 1. 
In the AJCM case, the reversible oscillation $\ket{1,g}\Leftrightarrow\ket{2,e}$ competes with the irreversible relaxation process $\ket{2,e}\Rightarrow\ket{0,e}$, and eventually transfer all population to the state \(\ket{0,e}\). 
Similar to the even parity case, we also observe a peak (valley) in the mean photon number displayed in Fig.~\ref{SteadyStateOddFig}(a) (the qubit population in Fig.~\ref{SteadyStateOddFig}(b)) at around \(g_2/\omega=0.42\) with very small \(g_1/\omega\). 
These local extrema result from the Rabi-type oscillation induced by avoided crossings, i.e.
\begin{equation}\label{OddSSoscillation}
    \ket{0,g}\leftrightarrow \ket{0,-}_\text{AJC},
\end{equation}
where \(\ket{0,-}_\text{AJC}\) is a superposition of the bare states \(\ket{0,g}\) and \(\ket{1,e}\), as given in Eq.~(\ref{AJCeigenstates}). 
With simple calculations, we know that the avoided crossings are present at
\begin{equation}
    \frac{g_2}{\omega} = \sqrt{(2m-1)-\frac{\Delta}{\omega}},\quad m\in\mathbb{N}_+,
\end{equation}
with small value of \(g_1\).

~

\section{Summary}\label{SectionSummary}
In summary, our investigation delves into the interplay among three types of interactions - the Jaynes-Cummings interaction, the counter-rotating terms, and the two-photon relaxation - within the framework of the anisotropic Rabi model.

To comprehensively explore these competitive dynamics, we employ three approaches: numerical solution of the master equation, interpretation of the transition diagram induced by the interaction terms, and understanding of the dynamics through the eigenbasis of the effective Hamiltonian.

We observe that the dynamics of cavity photon number and qubit population are sensitive to parity, which can be elucidated through the analysis of three solvable limits of the non-Hermitian effective Hamiltonian. 
Specifically, we identify that the intriguing local maximum in the transient state for small $g_1$ is absent in the steady state, and offer an intuitive explanation for its origin. 
Furthermore, we discover peaks in the steady states of cavity photon number and qubit population, attributed to avoided level crossings in the eigenvalue spectrum.

Our work contributes to a deeper understanding of the competition between Jaynes-Cummings interactions and counter-rotating terms, as well as the effects of ultrastrong coupling in open quantum systems. 
We provide clarification on the effects of the quantum jump operator and offer a generalized framework for exploring the interplay among different types of interactions. 
These findings may have implications for cavity-based quantum computations.

~

\begin{acknowledgments}
This work has been supported by the National Natural Science Foundation of China Grant No. 12205383, 12274470, 62273361 and the Natural Science Foundation of Changsha Grant No. kq2202082.
\end{acknowledgments}

~

\appendix

~

\section{Dynamics calculated from the non-Hermitian effective Hamiltonian}
We calculate the system dynamics by numerically solving the master equation in the main text. Here, we study the changes in the cavity photon number and qubit population of the system with the coupling strength after an extended period of evolution, utilizing the time-dependent Schr$\ddot{\rm o}$dinger equation(TDSE) derived from the effective Hamiltonian. In addition, by comparing the results obtained by the two approaches, we find the crucial role of the quantum jump term in reaching the ultimate steady state.

\begin{figure}
    \centering
    \includegraphics[width=\linewidth]{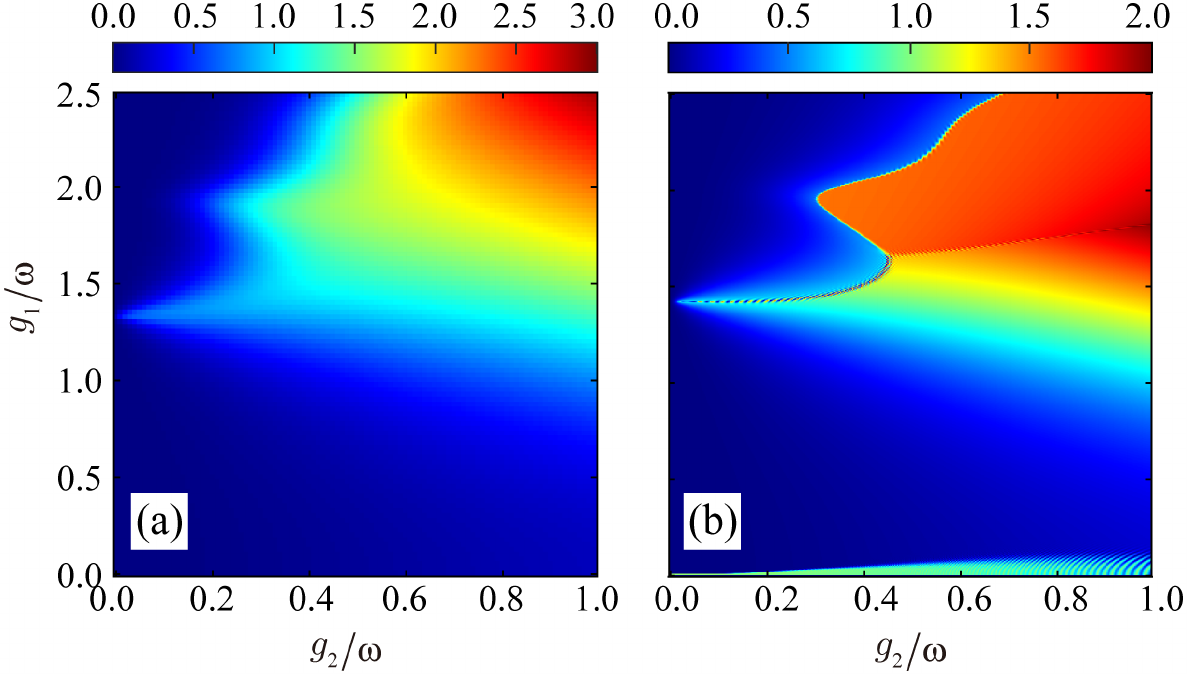}
    \caption{The cavity photon number of steady state in even parity subspace, calculated through (a) the Lindblad master equation and (b) the TDSE of the non-Hermitian effective Hamiltonian, with $\Delta/\omega=0.8$, $\kappa/\omega=0.02$. }
    \label{fig:Fig7}
\end{figure}

The general states of the system can be extended in the order of the basis vectors of the Hilbert space. In the even parity subspace, when the total photon number $N$ is even, the system's state can be represented as
\begin{equation}
    \ket{\psi(t)}=\sum_{n=0}^{N/2}C_{2n}^g(t)\ket{2n,g}+\sum_{n=0}^{N/2-1}C_{2n+1}^e(t)\ket{2n+1,e} ,
\end{equation}
with the probability amplitudes $C_{2n}^g(t)$, $C_{2n+1}^e(t)$, which can be obtained by solving the TDSE
\begin{equation}
    i\dot{\ket{\psi(t)}}=H_{\rm eff}\ket{\psi(t)}.
\end{equation}
with the initial state $\ket{\psi(0)}=\ket{2,g}$. As shown in \ref{fig:Fig7}, after a sufficiently long evolution time $(10^3\pi)$, we observe variations in $\langle a^{\dagger}a\rangle$ for different coupling strengths $g_1,g_2$ under the evolution dominated by the effective Hamiltonian. The positions of characteristic peaks align with those obtained by solving the master equation. However, some oscillations are present in these steady state results, indicating they are not fully stabilized but rather more stable states. 

The master equation approach represents the behavior of the true final steady state due to its consideration of all effects including the quantum jump term $\gamma\sigma_-\rho\sigma_+$, whereas the effective Hamiltonian omits it. Therefore, we can conclude that the inclusion of the quantum jump terms allows the system to reach the ultimate steady state.

~

\section{Physical interpretation of the Lindblad terms}
In this section, we discuss the three dissipation terms in the Lindblad master equation. Here, we consider two bare states in the basis of the main body system, denoted as $\ket{1}=\ket{n-2,g}$ and $\ket{2}=\ket{n,g}$, which are connected by the two-photon relaxation. Consequently, the density operator can be expressed as $\rho=\sum_{i,j=1,2}\rho_{ij}\ket{i}\bra{j}$ or matrix form
\begin{equation}
    \rho=\begin{pmatrix}
        \rho_{22}&\rho_{21}\\
        \rho_{12}&\rho_{11}
    \end{pmatrix},
\end{equation}
where the diagonal elements represent population $(\sum_i\rho_{i,i}=1$ with $\rho_{ii}\in\mathcal{R_0^+})$, while the off-diagonal elements denote coherence $(\rho_{ij}=\rho_{ji}^*$ with $\rho_{ij}\in\mathcal{C})$.

The dynamics of the system can be fully described by employing the Lindblad master equation, where the relaxation of photon pairs is represented by the following dissipator
\begin{equation}
    \mathcal{D}[a^2]\rho=2\kappa a^2\rho(a^{\dagger})^2-\kappa(a^{\dagger})^2a^2\rho-\kappa\rho(a^{\dagger})^2a^2.
\end{equation}
with relaxation rate $\kappa$. The effect of the dissipator on the density operator can be divided into two parts: the continuous nonunitary dissipation terms given by $\kappa\{(a^{\dagger})^2a^2,\rho\}$, and the quantum jump terms represented by $2\kappa a^2\rho(a^{\dagger})^2$. We can obtain a matrix form for the dissipator as
\begin{equation}
    \mathcal{D}[a^2]\rho=n(n-1)\begin{pmatrix}
        -2\kappa\rho_{22}&-\kappa\rho_{21}\\
        -\kappa\rho_{12}&2\kappa\rho_{22}
    \end{pmatrix}.
\end{equation}
By examining each term individually within this matrix, we intuitively reveal the effects of each element. Specifically speaking:

$\bullet$ The nonunitary dissipation terms impact elements such as $\rho_{21}$, $\rho_{12}$ along with coherence loss between bare states, which are represented by $-\kappa\rho_{21}$, $-\kappa\rho_{12}$ respectively.

$\bullet$ Additionally, the nonunitary dissipation terms also lead to energy and information losses from the upper state into the environment, related to $-2\kappa\rho_{22}$.

$\bullet$ On the contrary to the above-mentioned effects, the quantum jump terms contribute towards an increase in lower states' population, that is, $2\kappa\rho_{22}$.

Overall, the excited number of the two bare states is conserved over time i.e., $Tr([\mathcal{D}[a^2]\rho)=0$.

~

\section{Derivations of the JC and AJC results}
In this section, we study the eigensystems of the JCM and AJCM with two-photon relaxation, whose Hilbert space split into two unconnected parity chains due to the $Z_2$-symmetry. 

When $g_2=0$, the adjacent states in each parity subspace are solely coupled by rotating terms, resulting in Hilbert space partitioned into a collection of JC doublets $\{\ket{n, e},\ket{n+1, g}\}$. 
The effective Hamiltonian of the JCM with two-photon relaxation can be expressed in matrix form as
\begin{equation}
\begin{adjustbox}{max width=\linewidth}
    $
     H_{\rm eff}^{\rm JC}=\begin{pmatrix}
       (n+\frac{1}{2})\omega-\frac{\delta}{2}-i\kappa n(n-1)&g_1\sqrt{n+1}  \\
        g_1\sqrt{n+1} & (n+\frac{1}{2})\omega+\frac{\delta}{2}-i\kappa n(n+1)
    \end{pmatrix},
    $
\end{adjustbox}
\end{equation}
with $\delta=\omega-\Delta$ denoting the detuning. The complex eigenvalues of the block Hamiltonian are given by
\begin{equation}
    E_{n,\pm}^{\rm JC}=(n+\frac{1}{2})\omega-i\kappa n^2\pm\frac{1}{2}\Omega_n^{\rm JC},
\end{equation}
with Rabi frequency $\Omega_n^{\rm JC}=\sqrt{(\delta-2i\kappa n)^2+4g_1^2(n+1)}$. 
Meanwhile, the corresponding eigenstates are found as
\begin{align}\label{JCeigenstates}
    &\ket{n,+}_{\rm JC}=\cos\theta_n\ket{n,e}+\sin\theta_n\ket{n+1,g},\\
    &\ket{n,-}_{\rm JC}=-\sin\theta_n\ket{n,e}+\cos\theta_n\ket{n+1,g},
\end{align}
where the probability coefficients in the states are defined as
\begin{align}
    &\cos\theta_n=\sqrt{\frac{\Omega_n^{\rm JC}-\delta+2i\kappa n}{2\Omega_n^{\rm JC}}},\\
    &\sin\theta_n=\sqrt{\frac{\Omega_n^{\rm JC}+\delta-2i\kappa n}{2\Omega_n^{\rm JC}}}.
\end{align}

Because of the excitation number conserving with this system, this leads the Rabi oscillations between two bare states $\ket{n,e}$ and $\ket{n+1,g}$. 
In this way, we can assume the form of the time evolution of the system states as
\begin{equation}
\begin{adjustbox}{max width=\linewidth}
    $
    \ket{\psi(t)}=[C_n^e(t)\ket{n,e}+C_{n+1}^g(t)\ket{n+1,g}]e^{-i(n+1/2)\omega t}.
    $
    \end{adjustbox}
\end{equation}
where the $C_n^e(t)$ and $C_{n+1}^g(t)$ are time-dependent probability amplitudes for each state. 
Applying the TDSE $i\dot{\ket{\psi(t)}}=H_{\rm eff}^{\rm JC}\ket{\psi(t)}$, we then obtain the equations that the amplitudes satisified,
\begin{equation}
\begin{adjustbox}{max width=\linewidth}
$
\begin{aligned}
    &\dot{C}_n^e(t)=[i\frac{\delta}{2}-\kappa n(n-1)]C_n^e(t)-ig_1\sqrt{n+1}C_{n+1}^g(t),\\
    &\dot{C}_{n+1}^g(t)=[-i\frac{\delta}{2}-\kappa n(n+1)]C_{n+1}^g(t)-ig_1\sqrt{n+1}C_n^e(t).
\end{aligned}
$
\end{adjustbox}
\end{equation}
When the initial states meet $C_n^e(0)=1, C_{n+1}^g(0)=0$, and $\delta=0$, we can calculate the time evolution equation of the system state as follows
\begin{equation}
\begin{adjustbox}{max width=\linewidth}
    $
    \begin{aligned}
        \ket{\psi(t)}=&[(\cos\frac{\Omega_n^{\rm JC}t}{2}+\frac{2\kappa n}{\Omega_n^{\rm JC}}\sin\frac{\Omega_n^{\rm JC}t}{2})\ket{n,e}\\
        &-i\frac{2g_1\sqrt{n+1}}{\Omega_n^{\rm JC}}\sin\frac{\Omega_n^{\rm JC}t}{2}\ket{n+1,g}]e^{-i(n+1/2)\omega t-\kappa n^2 t}.
\end{aligned}
$
\end{adjustbox}
\end{equation}
We then obtain the time-dependent population on state $\ket{n+1,g}$ is
\begin{equation}
    P_{n+1,g}^{\rm JC}=\frac{4g_1^2(n+1)}{{\Omega_n^{\rm JC}}^2}\sin^2\frac{\Omega_n^{\rm JC}t}{2}.
\end{equation}

Similarly, when $g_1=0$, the ARM is reduced to AJCM where adjacent states within each parity subspace are solely coupled through counter-rotating terms, thus the Hilbert space is divided into blocks spanned by the basis $\{\ket{n+1,e},\ket{n,g}\}$. 
Consequently, the effective Hamiltonian of the AJCM with two-photon relaxation can be written in matrix form as
\begin{equation}
\begin{adjustbox}{max width=\linewidth}
$
H_{\rm eff}^{\rm AJCM}=\begin{pmatrix}
           (n+\frac{1}{2})\omega+\frac{\xi}{2}-i\kappa n(n+1) & g_2\sqrt{n+1} \\
        g_2\sqrt{n+1} & (n+\frac{1}{2})\omega-\frac{\xi}{2}-i\kappa n(n-1)
    \end{pmatrix},
$
\end{adjustbox}
\end{equation}
with $\xi=\omega+\Delta$. Then the complex eigenvalues of the partitioned matrix are given by
\begin{equation}
    E_{n,\pm}^{\rm AJC}=(n+\frac{1}{2})\omega-i\kappa n^2\pm\frac{1}{2}\Omega_n^{\rm AJC},
\end{equation}
with $\Omega_n^{\rm AJC}=\sqrt{(\xi-2i\kappa n)^2+4g_2^2(n+1)}$ denoting the Rabi frequency. 
The eigenstates can also be obtained 
\begin{align}\label{AJCeigenstates}
    &\ket{n,+}_{\rm AJC}=\cos\varphi_n\ket{n+1,e}+\sin\varphi_n\ket{n,g},\\
    &\ket{n,-}_{\rm AJC}=-\sin\varphi_n\ket{n+1,e}+\cos\varphi_n\ket{n,g},
\end{align}
and 
\begin{align}
    &\cos\varphi_n=\sqrt{\frac{\Omega_n^{\rm AJC}+\xi-2i\kappa n}{2\Omega_n^{\rm AJC}}},\\
    &\sin\varphi_n=\sqrt{\frac{\Omega_n^{\rm AJC}-\xi+2i\kappa n}{2\Omega_n^{\rm AJC}}}.
\end{align}

Because of the counter-rotating terms, when the initial state of the system is $\ket{n+1,e}$, only $\ket{g,n}$ state can be coupled with it. 
Then we can assume the form of the time evolution of the system states as
\begin{equation}
\begin{adjustbox}{max width=\linewidth}
    $
    \ket{\phi(t)}=[C^e_{n+1}(t)\ket{n+1,e}+C^g_n(t)\ket{g,n}]e^{-i(n+1/2)\omega t}.
    $
    \end{adjustbox}
\end{equation}
with the $C_{n+1}^e(t)$ and $C_n^g(t)$ denoting the time-dependent probability amplitudes for each state. 
Using the TDSE $i\dot{\ket{\phi(t)}}=H_{\rm eff}^{\rm AJC}\ket{\phi(t)}$, we then get the equations that the amplitudes satisfied,
\begin{equation}
\begin{adjustbox}{max width=\linewidth}
$
\begin{aligned}
    &\dot{C_{n+1}^e}(t)=[-i\frac{\xi}{2}-\kappa n(n+1)]C_{n+1}^e(t)-ig_2\sqrt{n+1}C_n^g(t),\\
    &\dot{C_n^g}(t)=[i\frac{\xi}{2}-\kappa n(n-1)]C_n^g(t)-ig_2\sqrt{n+1}C^e_{n+1}(t).
\end{aligned}
$
\end{adjustbox}
\end{equation}
When the initial states meet $C_{n+1}^e(0)=1, C_n^g(0)=0$, we then get the time evolution equation of the system state as follows
\begin{equation}
\begin{adjustbox}{max width=\linewidth}
    $
    \begin{aligned}
        \ket{\phi(t)}=&[(\cos\frac{\Omega_n^{\rm AJC}t}{2}-\frac{i\xi+2\kappa n}{\Omega_n^{AJC}}\sin\frac{\Omega_n^{\rm AJC}t}{2})\ket{n+1,e}\\
        &-i\frac{2g_2\sqrt{n+1}}{\Omega_n^{AJC}}\sin\frac{\Omega_n^{\rm AJC}t}{2}\ket{g,n}]e^{-i(n+1/2)\omega t-\kappa n^2 t}
    \end{aligned}
    $
    \end{adjustbox}
\end{equation}
We then obtain the time-dependent population on state $\ket{n,g}$ is
\begin{equation}
    P_{n,g}^{\rm AJC}=\frac{4g_2^2(n+1)}{{\Omega_n^{\rm AJC}}^2}\sin^2\frac{\Omega_n^{\rm AJC}t}{2}.
\end{equation}

~

\section{Further discussions on the parity-time-symmetry}\label{AppendixPT}
The energy spectrum of the JCM with two-photon dissipation is characterized by parity-time symmetry and the presence of exceptional points (EPs). Based on the above calculation, the eigenvalues of this system at resonate are
\begin{equation}
    E_{n,\pm}^{\rm JC}=(n+\frac{1}{2})\omega-i\kappa n^2\pm\sqrt{g^2(n+1)-\kappa^2n^2},
\end{equation}
where $-i\kappa n^2$ is interpreted as a frequency shift.  
For $g^2(n+1)>\kappa^2n^2$, the radical term has a real value, whereas for $g^2(n+1)<\kappa^2n^2$, the value becomes the following imaginary form
\begin{equation}
    E_{n,\pm}^{\rm JC}=(n+\frac{1}{2})\omega-i\kappa n^2\pm i\sqrt{\kappa^2n^2-g^2(n+1)}.
\end{equation}

\begin{figure}
    \centering
    \includegraphics[width=\linewidth]{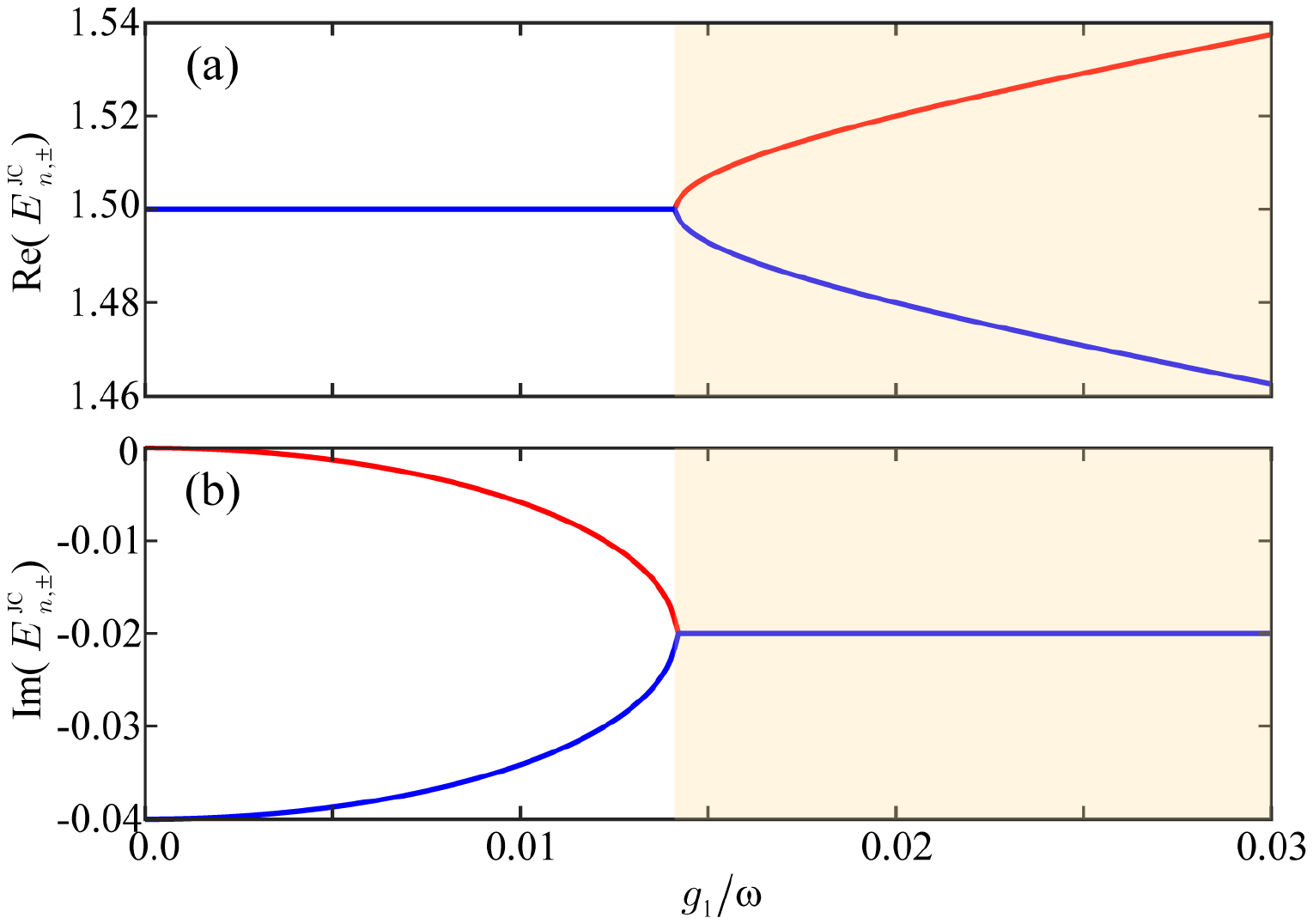}
    \caption{Energy spectrum of the JCM with two-photon dissipation when $\Delta=\omega$ and $n=1$. (a) and (b) describe the real and imaginary parts of the splitting energy levels $E_{1, +}^{\rm JC}$ (red line) and $E_{1, -}^{\rm JC}$ (blue line), respectively. It can be observed that there is an energy overlap at $g_1/\omega=\sqrt{2}\times10^{-2}$. }
    \label{fig:Fig8}
\end{figure}

According to the energy spectrum of the JC model with two-photon dissipation when $n=1$, the energy level splitting is closed at $g_1/\omega=\kappa/\sqrt{2}$ and the imaginary part of the energy after this position is a constant $-i\kappa$, which means the entire system has one dissipation channel at the same rate. In this case, it can be regarded as a passive $\mathcal{PT}$ system. The imaginary part of the energy level is zero after the crossing point by shifting the imaginary part of the energy level. The region of all-real eigenvalues corresponds to the $\mathcal{PT}$ symmetric phase of the system, as shown in the darker part of Fig.(\ref{fig:Fig8}).

Generally, the system reaches an exceptional point (EP) when $\delta=0$, and the position is
\begin{equation}
    g_1^{\rm EP}=\sqrt{\frac{\kappa^2n^2}{n+1}}.
\end{equation}
which varies with the photon number $n$.

\end{document}